\documentclass[aip,jap,superscriptaddress,noshowpacs,nobalancelastpage,preprint]{revtex4-1}

\usepackage[pdftex]{graphicx}
\usepackage{bm}
\usepackage{color}
\usepackage{textcase, natbib, url}
\usepackage[normalem]{ulem}
\usepackage[colorlinks=true, linkcolor=blue, urlcolor=black, citecolor=blue]{hyperref}
\usepackage{amsmath,amssymb}
\usepackage{mathtools}
\usepackage{tabularx}
\usepackage{lettrine}

\usepackage{natbib}
\usepackage{siunitx}

\definecolor{darkblue}{rgb}{0,0,0.5}
\definecolor{darkgreen}{rgb}{0,0.5,0}
\definecolor{darkred}{rgb}{.9,0,0}
\definecolor{purple}{rgb}{0.5,0,0.6}
\definecolor{orange}{rgb}{1,0.5,0}
\definecolor{grey}{rgb}{.6,.6,.6}
\definecolor{lightpink}{rgb}{1,0.7,0.75}
\definecolor{pink}{rgb}{1,0.4,0.58}
\definecolor{deeppink}{rgb}{1,0.08,0.58}

\begin{document}

\title{Unveiling the bosonic nature of an ultrashort few-electron pulse}

\author{Gregoire Roussely}
\thanks{These authors have contributed equally}
\affiliation{Univ. Grenoble Alpes, CNRS, Grenoble INP, Institut N\'eel, 38000 Grenoble, France}

\author{Everton Arrighi}
\thanks{These authors have contributed equally}
\affiliation{Univ. Grenoble Alpes, CNRS, Grenoble INP, Institut N\'eel, 38000 Grenoble, France}

\author{Giorgos Georgiou}
\affiliation{Univ. Grenoble Alpes, CNRS, Grenoble INP, Institut N\'eel, 38000 Grenoble, France}
\affiliation{Univ. Savoie Mont-Blanc, CNRS, IMEP-LAHC, 73370 Le Bourget du Lac, France}

\author{Shintaro Takada}
\affiliation{Univ. Grenoble Alpes, CNRS, Grenoble INP, Institut N\'eel, 38000 Grenoble, France}
\affiliation{National Institute of Advanced Industrial Science and Technology (AIST), National Metrology Institute of Japan (NMIJ), Tsukuba, Ibaraki 305-8563, Japan}

\author{Martin Schalk}
\affiliation{Univ. Grenoble Alpes, CNRS, Grenoble INP, Institut N\'eel, 38000 Grenoble, France}

\author{Matias Urdampilleta}
\affiliation{Univ. Grenoble Alpes, CNRS, Grenoble INP, Institut N\'eel, 38000 Grenoble, France}

\author{Arne Ludwig}
\affiliation{Lehrstuhl f\"ur Angewandte Festk\"orperphysik, Ruhr-Universit\"at Bochum, Universit\"atsstrasse 150, 44780 Bochum, Germany.}

\author{Andreas D. Wieck}
\affiliation{Lehrstuhl f\"ur Angewandte Festk\"orperphysik, Ruhr-Universit\"at Bochum, Universit\"atsstrasse 150, 44780 Bochum, Germany.}

\author{Pacome Armagnat}
\affiliation{Univ. Grenoble Alpes, CEA, INAC-Pheliqs, 38000 Grenoble, France}

\author{Thomas Kloss}
\affiliation{Univ. Grenoble Alpes, CEA, INAC-Pheliqs, 38000 Grenoble, France}

\author{Xavier Waintal}
\affiliation{Univ. Grenoble Alpes, CEA, INAC-Pheliqs, 38000 Grenoble, France}

\author{Tristan Meunier}
\affiliation{Univ. Grenoble Alpes, CNRS, Grenoble INP, Institut N\'eel, 38000 Grenoble, France}

\author{Christopher B\"auerle}
\email[Corresponding author: ]{christopher.bauerle@neel.cnrs.fr}
\affiliation{Univ. Grenoble Alpes, CNRS, Grenoble INP, Institut N\'eel, 38000 Grenoble, France}

\begin{abstract}
\section*{Abstract}
Quantum dynamics is very sensitive to dimensionality. 
While two-dimensional electronic systems form Fermi liquids, one-dimensional systems -- 
Tomonaga-Luttinger liquids -- are described by purely bosonic excitations, even though they 
are initially made of fermions. 
With the advent of coherent single-electron sources, the quantum dynamics of such a liquid is now accessible at the single-electron level.
Here, we report on time-of-flight measurements of ultrashort few-electron charge pulses injected into a quasi one-dimensional quantum conductor.  
By changing the confinement potential we can tune the system from the one-dimensional Tomonaga-Luttinger liquid limit to the multi-channel Fermi liquid and show that the plasmon velocity can be varied over almost an order of magnitude.
These results are in quantitative agreement with a parameter-free theory 
and demonstrate a powerful new probe for directly investigating real-time dynamics of fractionalisation phenomena in low-dimensional conductors.\\
\end{abstract}

\maketitle

\section*{Introduction}
\lettrine[lraise=0.3,findent=5pt]{\textbf{A}}{} fundamental difference between bosons and fermions is that the former can be described at the classical macroscopic level while the latter cannot. 
In particular, in an ultrafast quantum nano-electronics setup, the experimentalist controls the - bosonic - electromagnetic degrees of the system and aims at injecting a single  - fermionic - coherent electron in the system. 
This interplay between bosonic and fermionic statistics is a central feature in one-dimensional quantum systems as it provides a unique playground for the study of interaction effects~\cite{giamarchi_book, glazman-yacoby_nature_2010}.

The reduced dimensionality influences the interaction between particles and can lead to fascinating phenomena such as spin-charge separation \cite{yacoby_science_2005}, charge fractionalisation \cite{yacoby_nphys_2008} or Wigner crystallisation \cite{ilani_nphys_2013}. 
The low-energy collective bosonic excitations consist of charge and spin density waves that propagate at two different velocities. 
While the spin density is unaffected by the Coulomb interaction and propagates at Fermi velocity $v_{\rm F}$, the charge density is strongly renormalised by the interactions and propagates with the plasmon velocity $v_{\rm P}$, which is usually much faster than the Fermi velocity. 
Spin-charge separation has been experimentally probed in momentum resolved tunnelling experiments
between two quantum wires \cite{yacoby_science_2005} as well as tunnelling from a quantum wire into a two-dimensional electron gas \cite{ford_science_2009}. 
In addition to spin-charge separation, charge fractionalisation occurs in one-dimensional systems \cite{safi_book_1995,safi_prb_1995,lederer_prb_2000,degiovanni_prb_2013}. 
Injecting an electron into a one-dimensional system with momentum conservation, the charge decomposes into right and left moving charge excitations, as demonstrated 
in [\cite{yacoby_nphys_2008}].
Charge fractionalisation also occurs in a system of two coupled Tomonaga-Luttinger liquids.
There, an electronic excitation present in one of the two channels fractionalises into a fast charge mode and a slow neutral mode, which are the eigenmodes of the coupled system\cite{sukhorukov_prb_08}.  This charge fractionalisation has been recently observed in a chiral two-channel Tomonaga-Luttinger liquid in the integer quantum Hall regime\cite{bocquillon_ncom_2013,heiblum_prl_2014,fujisawa_nnano_2014,feve_ncom_2015,fujisawa_nphys_2017}.

Here, we study the most general case where the system can be tuned continuously from a clean one-channel Tomonaga-Luttinger liquid to a multi-channel Fermi liquid in a non-chiral system. We use time-resolved measurement techniques~\cite{ashoori_prb_1992,haug_prl_1996} to determine the time of flight \cite{kamata_prb_2010,kataoka_prl_2016,kumada_ncom_2013} of a single-electron voltage pulse and extract the collective charge excitation velocity.
Our detailed modelling of the electrostatics of the sample allows us to construct and understand the excitations of the system in a parameter-free theory.
We show that our self-consistent calculations capture well the results of the measurements, validating the construction of the bosonic collective modes from the fermionic degrees of freedom.


\section*{Results}
\subsection*{Measurement principle}
\begin{figure}
\centering
\includegraphics[width=12.5cm]{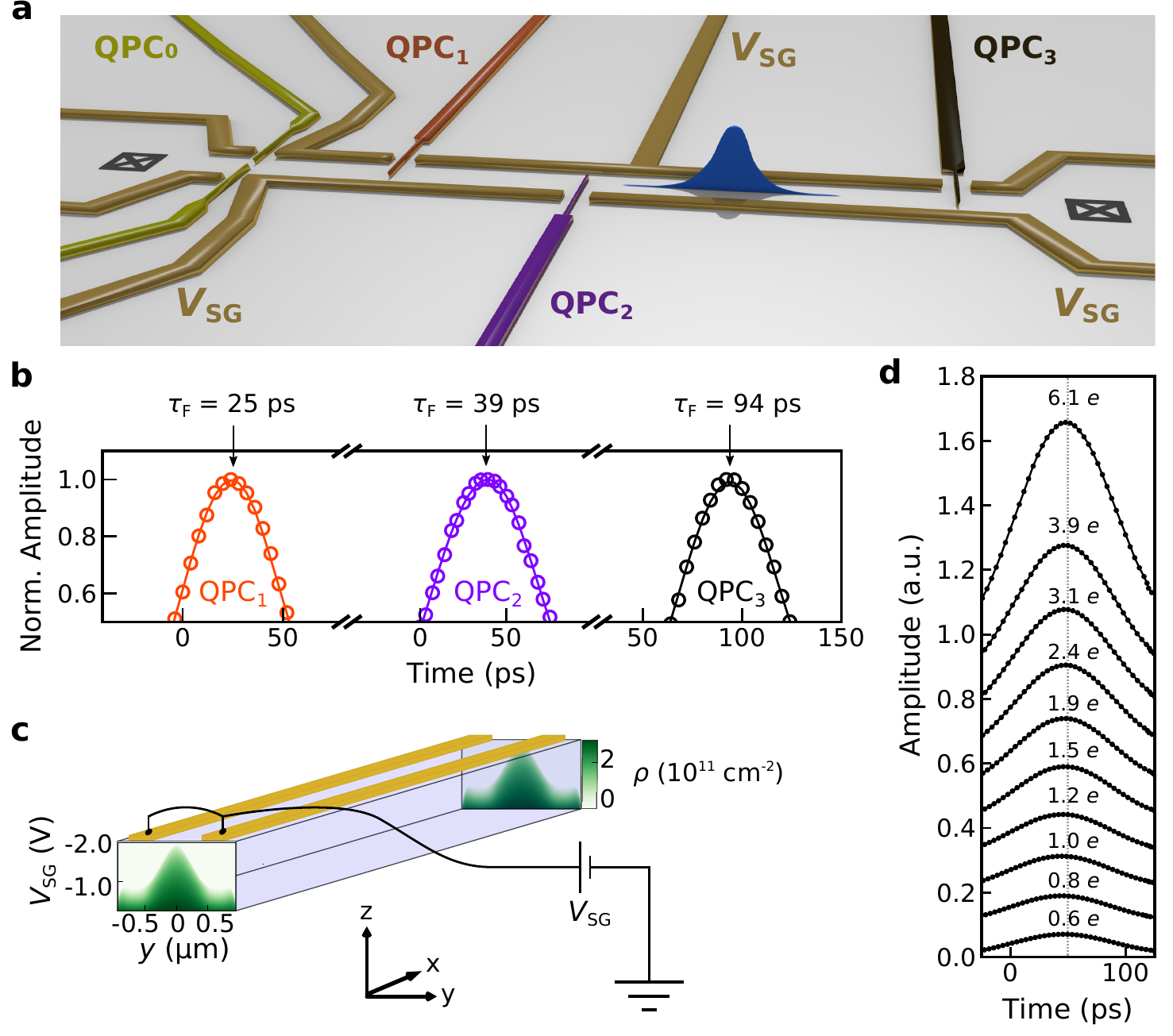}
\caption{\textbf{Device and time-of-flight measurements.} 
\textbf{a}, 
Schematic of  the quantum device. A few-electron pulse is launched at the left ohmic contact (black crossed box) by applying a very short ($\approx$ \SI{70}{\pico s}) voltage pulse. Three QPCs, denoted as QPC$_{1}$, QPC$_{2}$ and QPC$_{3}$ are placed along the quantum device at a distance of 15, 30 and 70 \si{\micro m} from the left ohmic contact. Each of these three QPCs is connected to a large bandwidth (\SI{40}{\giga \hertz}) bias tee and operated as an ultrafast switch. Time-resolved detection of current is done at the right ohmic contact.  QPC$_0$, placed a distance of \SI{6}{\micro m} from the left ohmic contact, is used as a channel selection.
\textbf{b}, Time-resolved measurements of an electron pulse at the three different QPC positions.
\textbf{c}, Illustration of the sample geometry used for the self consistent calculations. The quasi one-dimensional quantum wire is defined by the two long electrostatic gates at potential $V_{\rm{SG}}$. The colored images, one at the beginning of the wire and another one at the end, are cross sections of the electron density profile along the y-axis as a function of the gate voltage.
\textbf{d},  Time-resolved measurements of an electron pulse at QPC$_3$ for different excitation amplitudes. The amount of electrons contained in the electron pulse is varied between \SI{0.6}{\elementarycharge} and \SI{6.1}{\elementarycharge}.}
\label{fig:sample}
\end{figure}

We tailor a \SI{70}{\micro m} long quasi one-dimensional wire  into a two-dimensional electron gas using metallic surface gates as shown in Fig.~\ref{fig:sample}a. 
A pump-probe technique has been implemented to measure in a time-resolved manner the shape as well as the propagation speed of the electron pulse. 
We apply an ultrashort voltage pulse ($\approx$ \SI{70}{\pico s}) 
to the left ohmic contact to generate the few-electron pulse. 
The pulse injection is repeated at a frequency of \SI{600}{\mega \hertz} and the resulting DC current is measured at the right ohmic contact.
Three quantum point contacts (QPCs) are placed along the quantum wire to measure the arrival time of the charge pulse at different positions. 
Simultaneously, another ultrashort voltage pulse is sent to one of the three QPCs which allows opening and closing the QPC on a timescale much faster than the width of the few-electron pulse (see methods section).
By changing the time-delay between launching the electron pulse and the on--off switching of the QPC we can reconstruct the actual shape of the few-electron pulse \cite{kamata_prb_2010,kataoka_prl_2016}. 

\subsection*{Time-of-flight measurements} 
\begin{figure*}[ht!]
\centering
\includegraphics[width=0.9\linewidth]{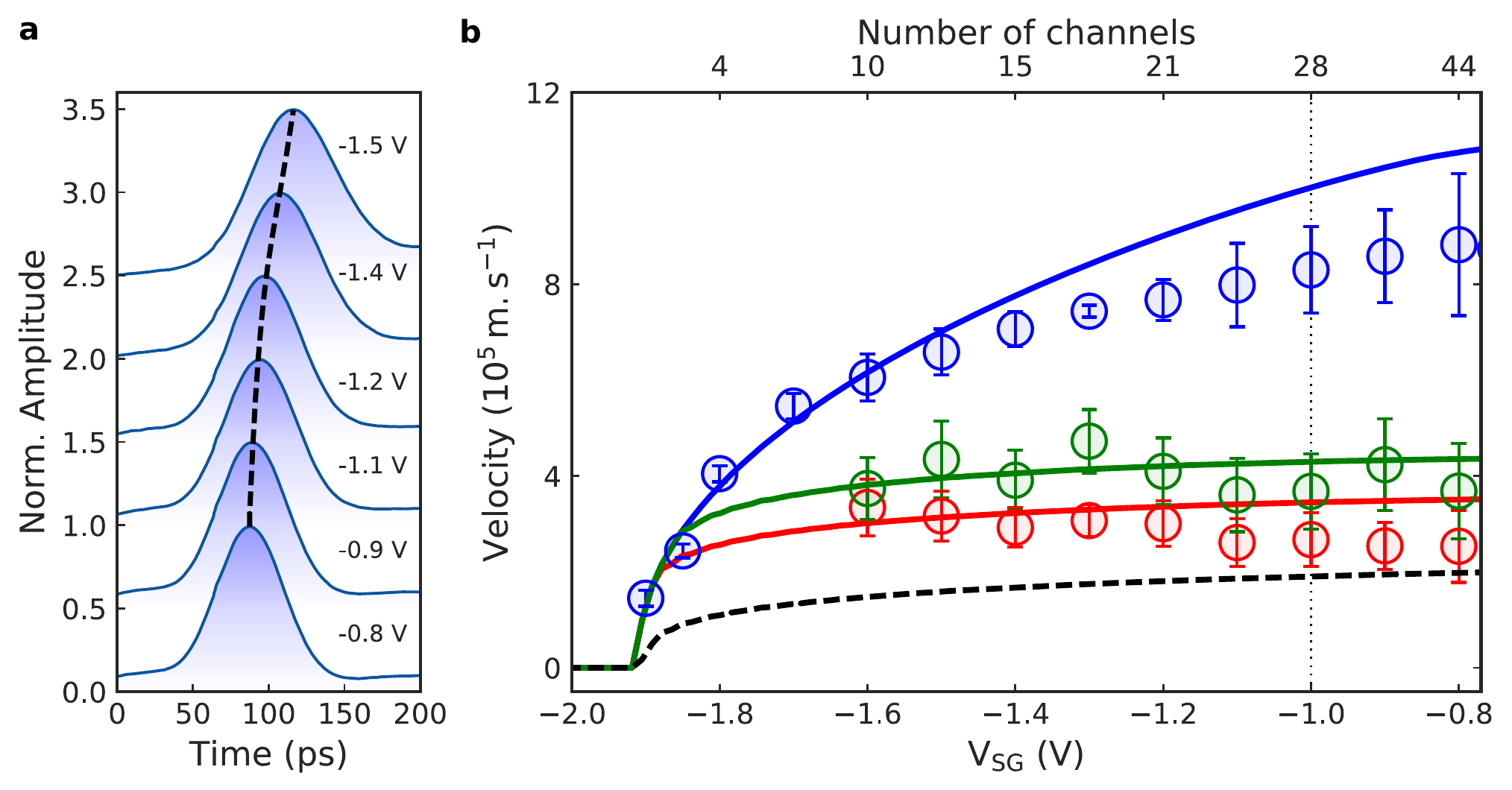}
\caption{\textbf{Tuning the propagation velocity.} 
\textbf{a}, Time-resolved measurements of the electron pulse for different confinement potentials at QPC$_3$ position. The curves have been offset vertically for clarity.
\textbf{b}, Velocity of the electron pulse as a function of the confinement potential and the corresponding number of channels. Open circles: experimental data, solid lines: parameter free self-consistent calculation. The blue data points correspond to the situation where the channel selection QPC$_0$ is not activated. The red (green) data are obtained by setting the channel selection QPC$_0$ at a conductance value of $G= 2e^2.h^{-1}$ ($G= 4e^2.h^{-1}$). The red (green) solid line corresponds to the velocity of the fast plasmon mode for a one-channel (two-channel) Tomonaga-Luttinger liquid. The black dashed line corresponds to the non-interacting Fermi liquid. The vertical dotted line indicates the gate voltage at which the velocities of Fig. \ref{fig:mode-selection}d are taken. The errors bars correspond to the velocity uncertainty and are derived from a linear fit of the QPC distance versus the respective time-of-flight (c.f. Supplementary Note 3).}
\label{fig:main}
\end{figure*}

A typical time-resolved measurement is shown in Fig. \ref{fig:sample}b.
We observe a few-electron pulse of Gaussian shape with a FWHM of $\approx$ \SI{70}{\pico s}. 
Measurements of the time of flight $\tau_{\rm F}$ at different positions (Fig. \ref{fig:sample}b) allows us to determine its propagation speed, which we find to be independent of the number of electrons contained in the electron pulse (Fig. \ref{fig:sample}d).
By changing the voltage on the side gates V$_{\rm{SG}}$ it is possible to modify the propagation speed by almost an order of magnitude. 
As the confinement is made stronger, the arrival time of the electron pulse at the detection QPC is shifted to longer times, as seen in Fig.\,\ref{fig:main}a.
This, as it will be demonstrated further on, is an indication of a slower propagation speed and it is in stark contrast to standard DC measurements. 
Indeed, in DC the Coulomb interaction is screened by the Fermi sea and the electrons travel at the Fermi velocity, as shown by magnetic focusing experiments \cite{focussing_delft}.
The situation is very different when creating a local perturbation of the charge density. 
Applying a very short charge pulse results in an excess charge density created locally. 
Due to the generated electric field, the excess charge is displaced very rapidly at the surface of the Fermi sea giving rise to a collective excitation, a plasmon \cite{chaplik_1985}.

\subsection*{Effect of Coulomb interaction on the propagation velocity} 

In one dimension, an interacting wire is described by Tomonaga-Luttinger plasmons of bosonic character\cite{giamarchi_book}.
The problem of generalising the bosonization construction to a system containing an arbitrary number of conduction channels, $N$, in the presence of Coulomb interactions has been treated theoretically by Matveev and Glazman\cite{matveev-glazman_1993}.  
The effect of the Coulomb potential is to couple the individual channels of the quantum wire, thus resulting in a collective behaviour that in turn affects strongly the propagation velocity of the excitations. 
For a quantum wire containing $N$ conduction channels, Coulomb interaction leads to charge fractionalisation into $N$ charge modes with renormalised propagation velocity and $N$ spin modes (c.f. Supplementary Note 4).
To distinguish between single-particle states and collective modes, we will use throughout the manuscript the term channel whenever referring to single-particle states and mode when referring to collective modes. As the spin modes do not carry any charge, their speed is not affected by the Coulomb interaction.
For our experiment we can neglect them since voltage pulses do not excite spin modes in the quantum conductor. The $N$ charge modes, on the other hand, are affected by the Coulomb interaction in the following way:
$N-1$ charge modes -- the slow modes -- are  weakly affected and propagate with a speed close to $v_{\rm F}$ while one mode -- the fast mode --  usually referred to as the plasmon mode is renormalised via all the other modes and propagates with a velocity much faster than $v_{\rm F}$.

Here, we have derived the theory \cite{matveev-glazman_1993} from first principles in order to obtain a quantitative - parameter free - comparison with the measurements. Our calculations proceed in three steps: first we solve the self-consistent electrostatics-quantum mechanics problem to obtain the effective potential seen by the electrons as shown in Fig.~\ref{fig:sample}c. Second, we compute the effective propagating channels and their interaction matrix and third we compute the mode velocities as arising from bosonisation theory (c.f. Supplementary Note 4).
The obtained theoretical data for the fast mode -- the plasmon mode -- (without any adjustable parameters) is displayed by the blue curve in Fig. \ref{fig:main}b. 

\begin{figure*} [ht!]
\centering
\includegraphics[width=0.8\linewidth]{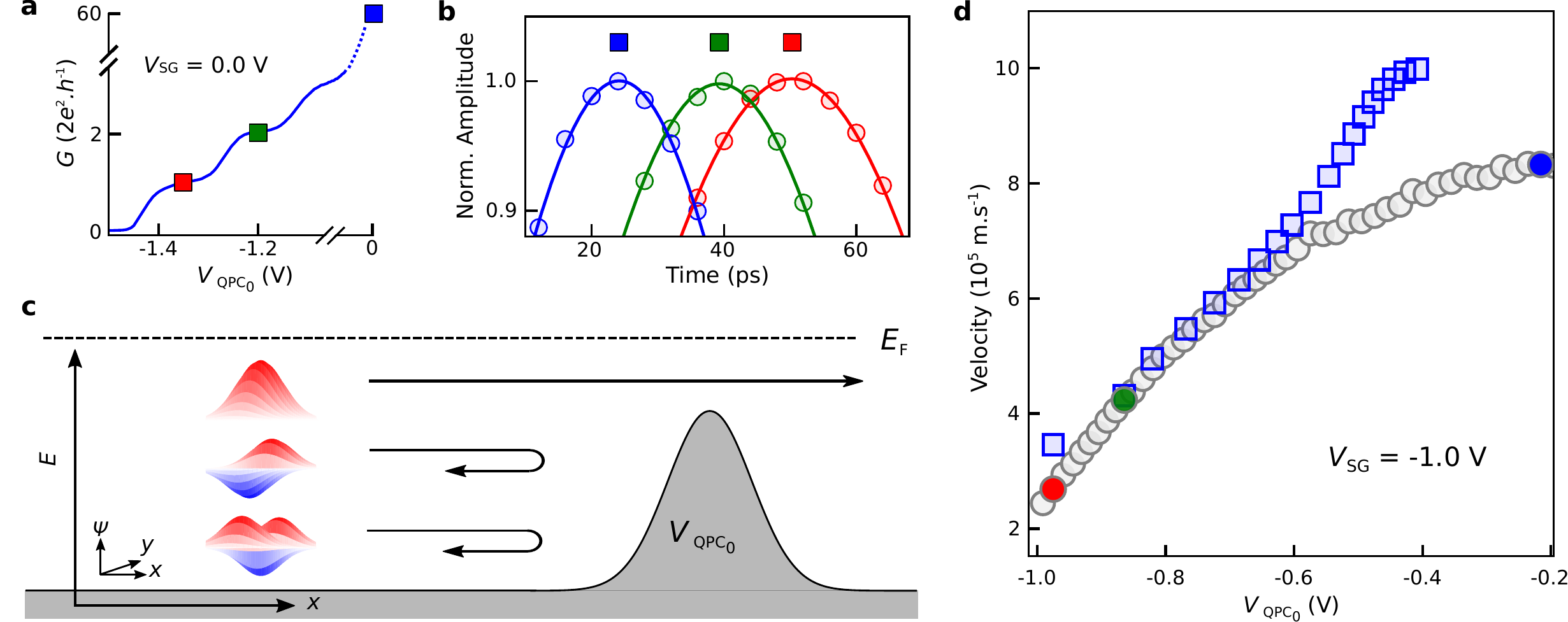}
\caption{\textbf{Channel selection of an electron pulse}. \textbf{a}, Conductance trace of the channel selection QPC$_0$ for $V_{\rm SG} = \SI{0.0}{\volt}$. \textbf{b}, Time-resolved measurements of the electron pulse propagation detected at QPC$_1$ position, when QPC$_0$ is deactivated (blue), or set to a conductance of $G= 2e^2.h^{-1}$ (red) and $G= 4e^2.h^{-1}$ (green). 
\textbf{c}, Schematic of the mode filtering experiment. 
The propagating voltage pulse populates all the available plasmon modes of the quantum wire (here for illustration purposes we show three) before passing through the channel selection QPC$_0$. 
The QPC$_0$, which is set to $G = 2e^2.h^{-1}$ reflects all channels except the one with the highest kinetic energy. After passing the channel selection QPC$_0$, only one single-channel plasmon mode is populated over a propagation distance of \SI{25}{\micro m}.
\textbf{d}, Propagation velocity of the electron pulse as a function of the channel selection QPC$_0$ voltage at a fixed confinement potential $V_{\rm{SG}} = \SI{-1.0}{\volt}$. The grey circles correspond to the experimentally measured velocity at QPC$_1$, while the blue squares is the outcome of a parameter-free calculation (c.f. Supplementary Note 4). The coloured circles correspond to the velocity measured for different conductance values of QPC$_0$, i.e. $G= 2e^2.h^{-1}$ (red circle), $G= 4e^2.h^{-1}$ (green circle) and fully depolarised QPC$_0$ (blue circle). These data points correspond to the dotted vertical line in Fig. \ref{fig:main}b at $V_{\rm SG} = \SI{-1.0}{\volt}$.
}

\label{fig:mode-selection}
\end{figure*} 

\subsection*{Channel selection}

By gradually reducing the number of channels of the quantum wire to one, we enter the Tomonaga-Luttinger liquid regime\cite{giamarchi_book}. 
However, due to the strong confinement potential and its long length, the quantum wire is not very homogeneous and the pulse becomes distorted. It is therefore not possible to realise a clean one-channel Tomonaga-Luttinger liquid\cite{pierre_ncom_2013} in the present configuration. 
To circumvent this limitation we have placed another quantum point contact QPC$_0$ at the entrance of the quantum wire in order to select specific channels as schematised in Fig. \ref{fig:mode-selection}c. 
We set the confinement potential of the quantum wire to a situation where the wire width is relatively large ($V_{\rm{SG}} = \SI{- 1.0}{\volt}$; $N \approx 28$) and set the quantum point conductance to a value of $G_{\rm{QPC_0}} = 2 e^2.h^{-1}$. 

An electron pulse is launched from the left ohmic contact into the quantum wire containing initially $N=28$ channels. Upon propagation, this charge pulse decomposes onto the $N=28$ eigenmodes (plasmon modes) due to Coulomb interaction. When this pulse passes through the QPC$_0$, only one channel is transmitted, as shown in Fig.\,\ref{fig:mode-selection}c. 
After the passage, the electron pulse continues its propagation along the quantum wire containing again the same number $N$ of available channels as before the passage through the selection QPC$_0$. 
Assuming a non-adiabatic passage, the charge pulse should instantaneously fractionalise into a fast plasmon mode and $N-1$ slow modes.
Very surprisingly, this is not the case. 
Time-resolved measurements of the charge pulse propagation through QPC$_1$, QPC$_2$ and QPC$_3$ allow us to determine the average speed of the charge pulse after passing through the selection QPC$_0$. 
We observe that the charge pulse is strongly slowed down after passing the channel selection QPC$_0$, as shown in Fig.\,\ref{fig:mode-selection}b and d. These measurements are repeated for different confinement potentials to corroborate our findings (see red data points in Fig.\,\ref{fig:main}b). 

\vspace{1cm}
\section*{Discussion}

As discussed above, the propagation speed of the charge pulse is strongly enhanced by the Coulomb interaction. 
Applying our parameter free model we are able to determine the propagation velocity for any gate configuration. This is done for the fast charge mode in Fig.\,\ref{fig:main}b (see blue continuous curve). 
The agreement with the experiments over the entire gate voltage region is quite remarkable. We attribute the observed discrepancy in the limit of large number of channels $N\sim 20-40$ to interchannel forward scattering which is not taken into account in \cite{matveev-glazman_1993}.

Our theoretical model also allows us to calculate the speed of the charge pulse assuming that only one single mode is occupied after passing the channel selection QPC$_0$ (solid red curve in Fig. \ref{fig:main}b) and compare it to our experimental data. 
This mode corresponds to a single-channel Tomanaga-Luttinger plasmon (c.f. Supplementary Note 4) which is very different from the plasmon hosted by the full 28 channels.
The agreement between theory and experiment is again remarkable. 
These observations strongly suggest that the charge pulse which is transmitted through the lowest channel of the selection QPC$_0$ is adiabatically transferred onto the fast plasmon mode corresponding to a  single-channel Tomonaga-Luttinger liquid and which we named the funneling scenario (c.f. Supplementary Note 4).
We have repeated these experiments for the second quantised plateau (green data points) and find similar agreement.
Hence our data indicate that it is  possible to form a very clean single channel (two-channel) Tomonaga-Luttinger liquid even though the wire contains many more active channels.
We observe that the electron pulse conserves its propagation speed for at least a distance of \SI{25}{\micro m} (position of QPC$_2$).
This is in stark contrast to experiments in the quantum Hall regime, where the wave packet fractionalises instantaneously\cite{feve_ncom_2015}.
In these experiments the electron wave packet is already fully fractionalised after a propagation distance of about \SI{3}{\micro m} \cite{feve_ncom_2015} with a time separation of $\approx$ \SI{70}{\pico s} between the  fast and the slow mode. 
In our experiment, we observe fractionalisation only at a distance well above \SI{20}{\micro m}. At a distance of about \SI{70}{\micro m} (QPC$_3$), we observe that the velocity is again approaching the one corresponding to the fast mode where all the channels of the quantum wire are populated. 
This opens the possibility to realise quantum interference experiments with single-electron pulses by only populating a single-channel plasmon mode, which has never been observed with DC measurements.

The presented time control of single-electron pulses at the picosecond level will also be important for the implementation of wave-guide architectures for flying qubits using single electrons\cite{bauerle_ropp_2018}.
Integrating a leviton source\cite{glattli_nature_2013} into a wave-guide interferometer would allow to realise single-electron flying qubit architectures\cite{yamamoto_nnano_2012,bautze_prb_2014,bauerle_ropp_2018} similar to those employed in linear quantum optics\cite{obrien_science_2008}.

Our findings give also a new insight into the recently discovered levitons\cite{glattli_nature_2013}.
As the underlying physics is independent of the actual shape of the single-electron wave packet, levitons should be regarded as a special kind of plasmon with the particularity that it does only generate electronic excitations (no holes), rather than a single-electron excitation propagating at the surface of the Fermi sea with the Fermi velocity\cite{levitov_jmp_1996}.

Furthermore, our studies pave the way for studying real-time dynamics of a quantum nano-electronic device \cite{gaury_nat-com_14} such as the measurement of the time spreading or the charge fractionalisation dynamics\cite{degiovanni_prb_2013} of the electron wave packet during propagation.

\newpage
\section*{Methods}
\subsection*{Sample fabrication}
\begin{figure}[h!]
\begin{center}
\includegraphics[width=12 cm]{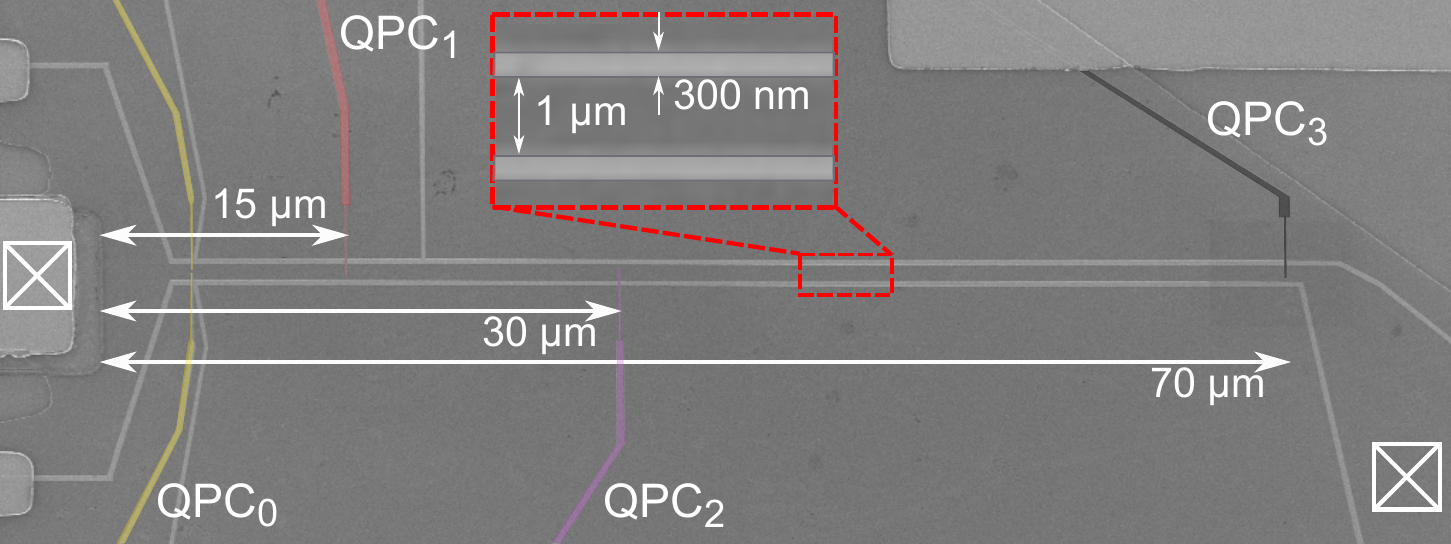}
\caption{\textbf{Scanning electron microscope image of the nanoelectronic device}. The light grey parts correspond to the electrostatic gates. The long quasi one-dimensional channel has length of \SI{70}{\micro m} and width of \SI{1}{\micro m}. The two ohmic contacts, one used for the excitation of the electron pulse (left) and the other one used for current detection (right) are indicated with crossed square boxes. The three QPC switches and their respective distance from the left ohmic contact are shown with red, pink and black colors, whereas the mode selection QPC$_0$ is highlighted with yellow color.}
\label{SEM}
\end{center}
\end{figure}

The sample is fabricated by depositing electrostatic gates on top of a GaAs/AlGaAs semiconductor heterostructure.
The two-dimensional electron gas, which is at a depth of \SI{140}{\nano m}, has density $n = \SI{2.11e11}{cm^{-2}}$ and mobility $\mu = \SI{1.89e6}{cm^{-2}V^{-1}s^{-1}}$, measured at \SI{4}{\kelvin}. 
The \SI{70}{\micro m} long electrostatic gates are defined by Ti/Au, while a Ni/Ge/Au/Ni/Au alloy is used for the ohmic contacts.
A scanning electron microscope image of our sample is shown in figure \ref{SEM}.\\ 
%

\subsection*{Time resolved measurements of voltage pulse}
\begin{figure}[h!]
\includegraphics[width=0.5\linewidth]{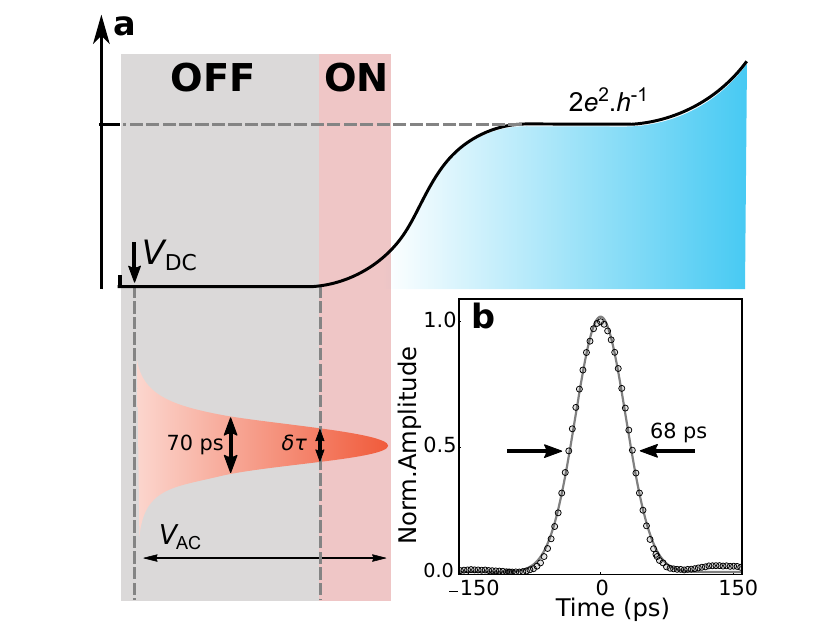}
\caption{\textbf{A QPC as a fast switch.} \textbf{a}, Operation diagram of our fast QPC switch. Initially we apply a negative DC voltage ($V_{\rm{DC}}$) on the QPC to set it deep into the pinch-off regime. Then we apply a fast pulse on the QPC, $V_{\rm{AC}}$, and gradually shift the DC voltage to higher values. By carefully choosing the right $V_{\rm{DC}}$ we can open the QPC for a very short time, $\sim$ \SI{10}{\pico s}.  \textbf{b}, Time resolved measurement of the electron pulse. The dark points are measurements and the continuous line is a Gaussian fit. The FWHM of the electron pulse is \SI{68}{\pico s}.}
\label{fig6:fastqpcswitch}
\end{figure}

To generate a single-electron pulse, a voltage pulse with an amplitude of several tens of \si{\micro \volt} is applied to the left ohmic contact of our sample through a high bandwidth coaxial line and a \SI{40}{\decibel} attenuation.
The voltage pulses are provided by an arbitrary function generator (Textronics AWG7122C) and have a \SI{600}{\mega \hertz} repetition frequency.
The generated DC current is measured across a \SI{10}{\kilo \ohm} resistor placed on the sample chip carrier at a temperature of \SI{20}{\milli \kelvin}. The pulse train is modulated at a frequency of \SI{12}{\kilo \hertz} to perform lock-in measurements.
A second voltage pulse is applied to one of the QPCs in order to operate it as a fast switch.
By changing the time delay between generating the electron pulse and opening/closing the QPC switch we can reconstruct in a time-resolved manner the time trace of the electron pulse, following the protocol developed by Kamata \textit{et al.} \cite{kamata_prb_2010}.\\
In order to obtain the shortest possible switching times we perform the following operations, shown in Fig.~\ref{fig6:fastqpcswitch}a.
First, the QPC is set to the pinch-off regime (OFF - position) by applying an appropriate negative DC voltage ($V_{\rm{DC}}$).
Subsequently, we apply a short voltage pulse with a fixed amplitude ($V_{\rm{AC}}$) to the QPC, which allows us to open the QPC switch only for a very short time $\delta\tau$, typically below \SI{10}{\pico s}\cite{johnson_apl_2017}. To achieve these fast switching times we keep the $V_{\rm{AC}}$ amplitude constant and we vary $V_{\rm{DC}}$. As shown in Fig.~\ref{fig6:fastqpcswitch}a when $V_{\rm{DC}}$ is very negative the QPC switch remains closed for all time delays and therefore the recorded current is zero. By increasing $V_{\rm{DC}}$ to the appropriate value we can open the QPC switch for a brief period of time $\delta\tau$, thus allowing us to reconstruct the electron pulse. 
As the switching profile of the QPC depends on the combination of the applied DC and AC voltages as well as the very sharp conductance response we can achieve time resolutions that are shorter than those provided by our electronics.
By optimising $V_{\rm{DC}}$ and $V_{\rm{AC}}$ amplitudes we are able to measure single-electron pulses down to a FWHM of \SI{68}{\pico s}, as shown in Fig. \ref{fig6:fastqpcswitch}b.\\

\subsection*{Determination of the propagation velocity}

To determine the velocity of the electron wave packet we perfomed time-of-flight measurements for different confinement potentials (see Fig.~\ref{fig:sample} and Fig.~\ref{fig:main}). For every confinement potential we carry out three independent measurements, one for each QPC (except QPC which is not connected to a bias tee).
During these measurements we excite the electron wave packet and measure the time it takes to propagate to the three detection QPCs. By using the time of flight and the exact distance between the left ohmic contact (excitation location) and these three QPCs (Fig. \ref{SEM}) we can calculate the velocity (cf. Supplementary Note 4).\\
%

\subsection*{Data availability:}The data that support the findings of this study are available from the corresponding authors on reasonable request.


\newpage
\section*{Acknowledgments}
We would like to dedicate this article to the late Frank Hekking who has been very implicated in the initial stage of the project. We would like to acknowledge fruitful discussions with  D. Carpentier, P. Degiovanni, M. Filippone and T. Martin. S. T. acknowledges financial support from the European Union's Horizon 2020 research and innovation program under the Marie Sk\l{}odowska-Curie grant agreement No. 654603. A. L. and A. D. W. acknowledge gratefully support of DFG-TRR160, BMBF - Q.com-H 16KIS0109, and the DFH/UFA CDFA-05-06. 
C. B, T.M. and X.W. acknowledge financial support from the French National Agency (ANR) in the frame of its program ANR Fully Quantum Project No. ANR-16-CE30-0015-02 and QTERA No. ANR-15-CE24-0007-02. C.B. and T.M. also acknowledge the SingleEIX Project No. ANR-15-CE24-0035. X.W. and P.A. are funded by the U.S. Office of Naval Research.

\section*{Author contribution}
G.R and E.A. performed the experiment and analysed the data with input from S.T., G.G, M.U. and T.M.. G.R made the sample with help from S.T. and T.M.. M.S. contributed to the experimental setup and timing methods. T.K., P.A. and X.W. designed the theoretical framework. P.A. performed the numerical simulations. All authors analyzed the numerical data. A.L. and A.D.W. provided the high mobility heterostructure. C.B, X.W. and G.G. wrote the paper with inputs from all authors. C.B. supervised the experimental work.

\textbf{Supplementary Information} accompanies this paper. 

\textbf{Competing interests:} The authors declare no competing financial or non-financial interests.
\end{document}


\title{Supplementary Information: Unveiling the bosonic nature of an ultrashort few-electron pulse}

\author{Gregoire Roussely$^\dagger$}
\affiliation{Univ. Grenoble Alpes, CNRS, Grenoble INP, Institut N\'eel, 38000 Grenoble, France}

\author{Everton Arrighi$^\dagger$}
\affiliation{Univ. Grenoble Alpes, CNRS, Grenoble INP, Institut N\'eel, 38000 Grenoble, France}

\author{Giorgos Georgiou}
\affiliation{Univ. Grenoble Alpes, CNRS, Grenoble INP, Institut N\'eel, 38000 Grenoble, France}
\affiliation{Univ. Savoie Mont-Blanc, CNRS, IMEP-LAHC, 73370 Le Bourget du Lac, France}

\author{Shintaro Takada}
\affiliation{Univ. Grenoble Alpes, CNRS, Grenoble INP, Institut N\'eel, 38000 Grenoble, France}
\affiliation{National Institute of Advanced Industrial Science and Technology (AIST), National Metrology Institute of Japan (NMIJ), Tsukuba, Ibaraki 305-8563, Japan}

\author{Martin Schalk}
\affiliation{Univ. Grenoble Alpes, CNRS, Grenoble INP, Institut N\'eel, 38000 Grenoble, France}

\author{Matias Urdampilleta}
\affiliation{Univ. Grenoble Alpes, CNRS, Grenoble INP, Institut N\'eel, 38000 Grenoble, France}

\author{Arne Ludwig}
\affiliation{Lehrstuhl f\"ur Angewandte Festk\"orperphysik, Ruhr-Universit\"at Bochum, Universit\"atsstrasse 150, 44780 Bochum, Germany.}

\author{Andreas D. Wieck}
\affiliation{Lehrstuhl f\"ur Angewandte Festk\"orperphysik, Ruhr-Universit\"at Bochum, Universit\"atsstrasse 150, 44780 Bochum, Germany.}

\author{Pacome Armagnat}
\affiliation{Univ. Grenoble Alpes, CEA, INAC-Pheliqs, 38000 Grenoble, France}

\author{Thomas Kloss}
\affiliation{Univ. Grenoble Alpes, CEA, INAC-Pheliqs, 38000 Grenoble, France}

\author{Xavier Waintal}
\affiliation{Univ. Grenoble Alpes, CEA, INAC-Pheliqs, 38000 Grenoble, France}

\author{Tristan Meunier}
\affiliation{Univ. Grenoble Alpes, CNRS, Grenoble INP, Institut N\'eel, 38000 Grenoble, France}

\author{Christopher B\"auerle}
\affiliation{Univ. Grenoble Alpes, CNRS, Grenoble INP, Institut N\'eel, 38000 Grenoble, France}

\maketitle

\noindent\section*{Supplementary Note 1}
\noindent\subsection*{Number of injected charges}
\noindent The number of electrons injected into the quasi-1D channel can be arbitrarily tuned by the voltage applied on the left ohmic contact, $V_{\rm{p}}$ \cite{gaury_nat-com_14},
\begin{equation*}
	n_e = 2 \frac{e}{h} \int V_{\rm{p}}(t) \ dt,
\end{equation*}
where $n_e$ is the number of excited electrons, $h$ is Plank's constant and $e$ is the elementary charge. In this formula we assume a single channel of conductance as well as spin degeneracy.
The voltage on the left ohmic contact, $V_{\rm{p}}(t)$, can be calculated directly from the voltage amplitude applied by the AWG on the RF line and by taking into account the appropriate line attenuation.

To estimate the number of generated electrons contained in one electron pulse we measure the rectified current across a 10~$\rm{k\Omega}$ resistor, which is amplified and measured with a lock-in amplifier. The measured signal is then proportional to the number of generated electrons $\overline{n}_{e}$ distributed over all available conduction channels and to the repetition rate of our AWG, which is $f=600~{\rm MHz}$, i.e.,
\begin{equation*}
   I = \overline{n}_{e}~e~f=\frac{V_{\rm{rms}}}{g}~\frac{1}{10~{\rm k \Omega}}~\frac{\pi}{\sqrt{2}}
\end{equation*}
where $V_{\rm{rms}}$ is the voltage measured with the lock-in amplifier, $g=1000$ is the amplifier gain and the constant factor $\pi/\sqrt{2}$ is to account for the $12~\rm{kHz}$ square pulse modulation signal used for the lock-in detection.
As shown in Fig.~1d of the manuscript we can excite from 0.6 to 6.1 electrons per pulse by varying the amplitude of the input pulse.

\newpage
\noindent\section*{Supplementary Note 2}
\noindent\subsection*{Time calibration of RF lines}
\noindent The RF lines used for the excitation and detection of the electron pulse are calibrated using two methods, reflectometry measurements and in-situ calibration. 

To perform the reflectometry calibration we initially send a short pulse through the RF line and measure the reflected pulse with a high bandwidth oscilloscope.
As shown in Supplementary Figure ~\ref{fig:reflectometry} we can estimate the relative time delay between the four RF lines by looking at the reflected pulses. This approach offers limited time accuracy ($\approx \pm$ 5 ps) since the RF lines cannot be calibrated with the attenuators installed. A second and more precise calibration can be done in-situ by exploiting the fast propagation of the 2D plasmon.

\begin{figure}[h!]
\includegraphics[width=0.7\linewidth]{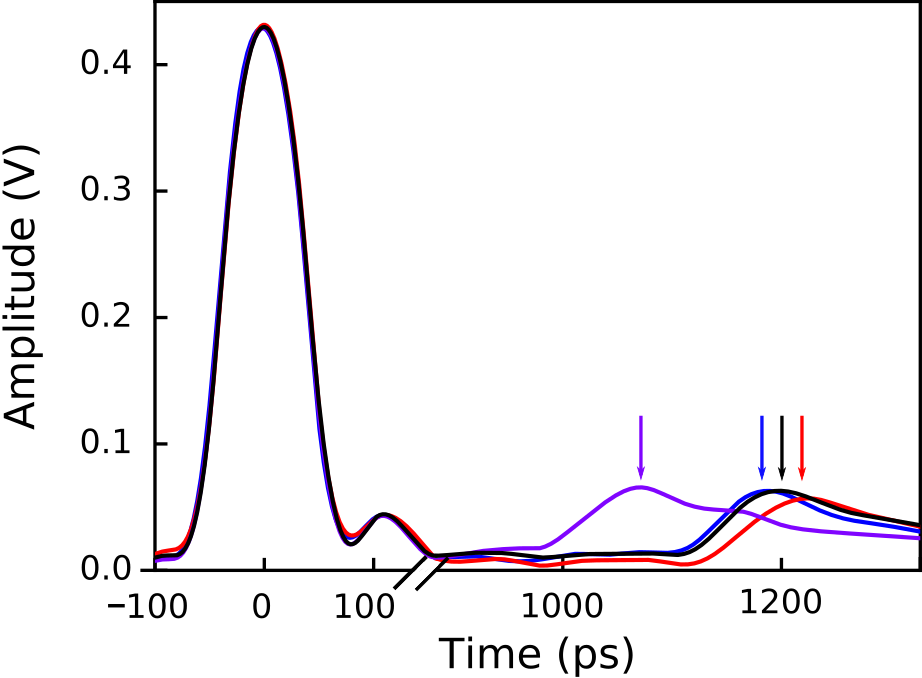}
\caption{\textbf{Time domain reflectometry}. The large amplitude curves at \SI{0}{\pico s} are the pulses applied on the RF lines, while the smaller amplitude curves above \SI{1000}{ps} are the reflected pulses. The four arrows indicate the peak of the reflected pulse connected to the ohmic contact (blue curve), QPC$_1$ (red curve), QPC$_2$ (purple curve), QPC$_3$ (black curve).}
\label{fig:reflectometry}
\end{figure}

To do so, we almost completely depolarise the one-dimensional channel gates. The collective motion of electrons in this almost confinement-free system has the velocity of a two-dimensional plasmon, which is well known to be $v_{\rm{plasmon}}^{\rm{2D}} \sim \SI{e7}{m.s^{-1}}\,$  \cite{chaplik_1985,2D_plasmon}. 
When the two-dimensional plasmon is injected at the left ohmic contact it will reach the sampling QPC essentially instantaneously. By measuring a time resolved trace of the two-dimensional plasmon pulse we can estimate the relative time delay between the excitation RF line and the detection QPC RF lines.
This allows us to calibrate our RF lines at low temperatures with an accuracy in the order of \SI{2}{\pico s}.

\newpage
\section*{Supplementary Note 3}
\noindent\subsection*{Time-of-flight measurements -- Velocity calculation}
\noindent The velocities shown in Fig.~2b of the manuscript (blue data points) are derived from time-of-flight measurements. As it was briefly explained in the methods section, to determine the velocities we performed three independent measurements, one for each QPC (except QPC$_0$ which is not connected to a bias tee).
For each measurement we monitored the time of flight of the electron wavepacket from the left ohmic contact to the respective QPC, resulting into $t_1$, $t_2$ and $t_3$ for QPC$_1$, QPC$_2$ and QPC$_3$ respectively.
The distance of the three QPCs from the left ohmic contact are derived from the Scanning Electron Microscope (SEM) image of our sample, shown in Fig.~4, and they are $d_1 = \SI{15}{\micro m}$, $d_2 = \SI{30}{\micro m}$ and $d_3 = \SI{70}{\micro m}$.  

\begin{figure}[h!]
\begin{center}
\includegraphics[width=0.7\linewidth]{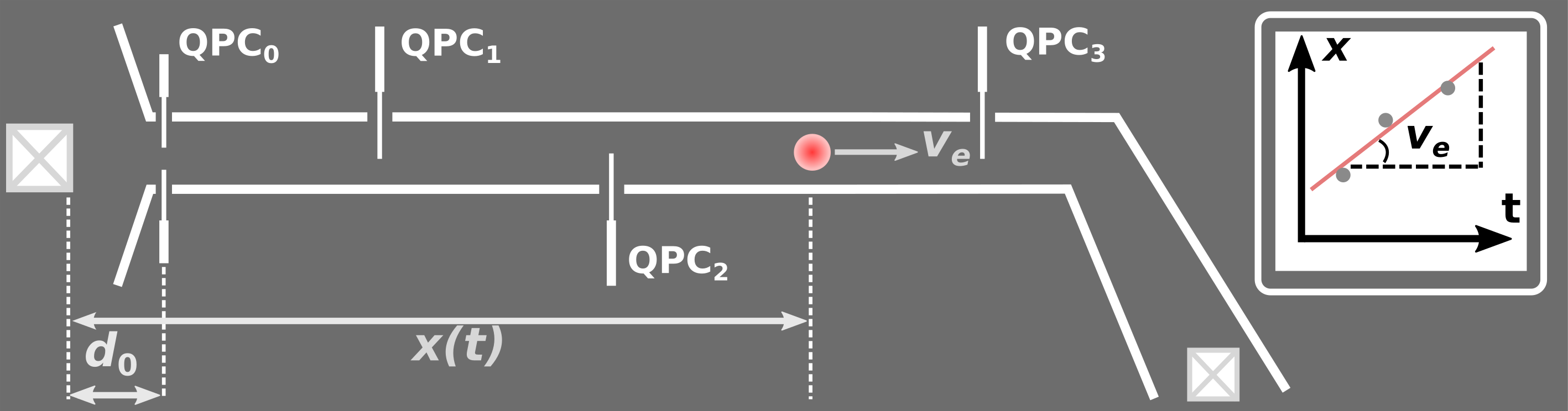}
\caption{\textbf{Velocity calculation}. Schematic illustrating the velocity calculation from the time-of-flight measurements. The electron wave packet (red sphere) is excited at the left ohmic contact (white crossed square) and it travels a distance $d_0$ before passing through the mode selection QPC$_0$ and entering the quasi one-dimensional wire. The wave packet will then travel at a speed $v_{\rm e}$ towards the three detection QPCs where a time-of-flight measurement is performed. The inset shows how the speed is derived from the three independent measurements.}
\label{fig:velocity_calculation}
\end{center}
\end{figure}

\noindent As illustrated by Supplementary Figure ~\ref{fig:velocity_calculation}, the electron wave packet excited at the left ohmic contact will cover a distance $d_{\rm{0}}$ with an average speed $v_{\rm{0}}$ before it reaches the quasi-1D channel. It will then travel towards the three QPCs with speed $v_{e}$.
The equation governing the motion of the electron has the following form,
\begin{equation}
   x = v_{\rm{0}} t_{\rm{0}} + v_{e} ( t - t_{\rm{0}} ) \\
   	 = d_{\rm{0}} + v_{e} t
     \label{linfit}
\end{equation}
where $x$ is the distance from the left ohmic contact to the three QPCs, $t_0$ is the time until the electron wave packet reaches the quasi-1D channel and $t$ is the time of flight.
The velocity can be calculated from the slope of Supplementary Eq. (\ref{linfit}) through a linear fit, as indicated by the inset of Supplementary Figure \ref{fig:velocity_calculation}.
The error bars on figure 2b of the main manuscript correspond to the velocity uncertainty obtained by the linear fit.

\newpage
\noindent\section*{Supplementary Note 4}
\noindent\subsection*{Simulating the plasmon dispersion relation from the microscopic model}
\noindent A sketch of the system used in our simulations is shown in Figure 1c of the main text.
We consider a 3D system with translational invariance along the x direction and with the 2D electron gas situated at $z=0$.
The two top gates, situated at $z=$ 140\,nm are used for defining the quasi-one dimensional wire, but also provide screening to the electron gas.

Our starting point is a many-body Hamiltonian that describes our 2D electron gas,
\begin{equation}
\label{eq:base}
H = -\frac{\hbar^2}{2m^*}\sum_\sigma \int d^2 \textbf{r} \ c^\dagger_{\textbf{r} \sigma} \Delta c^{\ }_{\textbf{r} \sigma}
+ \sum_\sigma \int d^2 \textbf{r} \ U(\textbf{r}) c^\dagger_{\textbf{r} \sigma} c^{\ }_{\textbf{r} \sigma}
+ \sum_{\sigma\sigma'}  \int d^2\textbf{r}  d^2\textbf{r}' c^\dagger_{\textbf{r} \sigma} c^{\ }_{\textbf{r}\sigma} G(\textbf{r},\textbf{r}')
c^\dagger_{\textbf{r}'\sigma'} c^{\ }_{\textbf{r}' \sigma'}
\end{equation}
where the fermionic operator $c^\dagger_{\textbf{r}\sigma}$ ($c^{\ }_{\textbf{r} \sigma}$) creates (destroys) an electron at position $\textbf{r} = (x,y)$ and with spin $\sigma$,
$m^*$ is the effective mass, $\Delta$ the 2D Laplacian operator, $U(\textbf{r})$ an electrostatic potential and $G(\textbf{r},\textbf{r}')$ the electron-electron interaction.
In free space, $G(\textbf{r},\textbf{r}')$ is simply given by the bare Coulomb repulsion $G(\textbf{r},\textbf{r}')=\frac{e^2}{4\pi\epsilon|\textbf{r}- \textbf{r}'|}$. However, here, the presence of the electrostatic gates provides some screening, and $G(\textbf{r},\textbf{r}')$ is the solution of the 3D Poisson equation (restricted to the 2D gas),
\begin{equation}
\Delta_{3D} G(\bf \hat r,\hat r') = - \it \frac{e^\textrm{2}}{\epsilon} \delta (\bf \hat r-\hat r')
\end{equation}
with the boundary condition $G(\bf \hat r,\hat r')= \rm 0$ when the 3D vector $\bf \hat r \it =(x,y,z)$ coincides with the position of a gate. $e>0$ is the electron charge and $\epsilon$ the dielectric constant.

\subsection*{Self-consistent electrostatic-quantum problem}
\noindent The first step in our calculation is a mean field (self-consistent Hartree) treatment of Supplementary Eq. (\ref{eq:base}), in which we aim at
solving together the following equations,

\begin{align}
		&\Delta_{3D} U(\bf \hat r) = -\it \frac{e \rho(\bf \hat r)}{\epsilon}+\it \frac{e \rho_{\textrm{0}}(\bf \hat r)}{\epsilon} \label{equ:real_poisson} \\
		&\frac{- \hbar^2}{2 m^\star} \Delta \Psi(\textbf{r}) - e U(\textbf{r}) \Psi(\textbf{r}) = E \Psi(\textbf{r}) \label{equ:real_schrodinger} \\
		&\rho(\textbf{r}) = \sum_E  f(E) |\Psi (E,\textbf{r})|^2  \label{equ:real_density}
\end{align}



where $U(\textbf{r})$ is the restriction of $U(\bf \hat r)$ to the 2D plane of the electron gas, $\rho(\bf \hat r)=\rho(r)\delta(\rm z)$, $\Psi$ the electronic wave function, $f$ the Fermi function and the continuum sum in Supplementary Eq. (\ref{equ:real_density}) spans over all the eigenstates of the Schr\"{o}dinger Supplementary Eq. (\ref{equ:real_schrodinger}). The density $\rho_0$ accounts for the layer of dopants present above the 2D gas; we use $U(\bf \hat r)= \it V_{\rm SG}$ in the top gate and Von Neumann boundary conditions otherwise. We have explicitly verified that the finite width of the 2D gas along $z$ does not play a role in these calculations. Translational invariance along $x$ implies that the Poisson equation can be solved in the 2D $(y,z)$ plane while the wave-function is a plane wave, which can be separated into a transverse and a longitudinal component
$\Psi (E,\textbf{r}) = e^{ik_{\alpha}(E)x} \psi_\alpha(y)$. Performing an explicit integration along the longitudinal direction at zero temperature, we arrive at, 

\begin{align}
		&\Delta U(y,z) = -\frac{e \rho(y)}{\epsilon} \delta (z) + \frac{e \rho_0(y,z)}{\epsilon} \label{equ:2D_poisson} \\
		&-\frac{\hbar^2}{2 m^\star} \frac{\partial^2}{\partial y^2} \psi_\alpha(y) - e U(y,0)  \psi_\alpha(y) = E_\alpha \psi_\alpha(y) 	\label{equ:1D_schrodinger}\\
		&\rho(y)=  \frac{2}{ \pi}\sqrt{\frac{2 m^\star}{\hbar^2}}  \sum_{\alpha = 0}^{N-1} |\psi_\alpha(y)|^2 \, \sqrt{E_{\textrm{F}}-E_\alpha}  \label{equ:2D_density}
\end{align}

%
where we have introduced the Fermi energy $E_{\rm F}$. The factor 2 accounts for spin degeneracy.
The Poisson equation for the potential $U(y,z)$  is solved on the $(y,z)$ plane by using finite elements in a 
rectangular box with Von Neumann boundary conditions on the sides and fixed Dirichlet boundary conditions for the side-gate voltage $V_{\rm{SG}}$. The size of the box has been chosen such that the results are free from finite size effects.
An example of calculation of the Green's function $G(y,y')\equiv G(y,z=0;y',z'=0)$ is shown in Supplementary  Fig. \ref{fig:green_function} 
for illustration.
\begin{figure}[h!]
	\centerline{\includegraphics[width=0.6\linewidth]{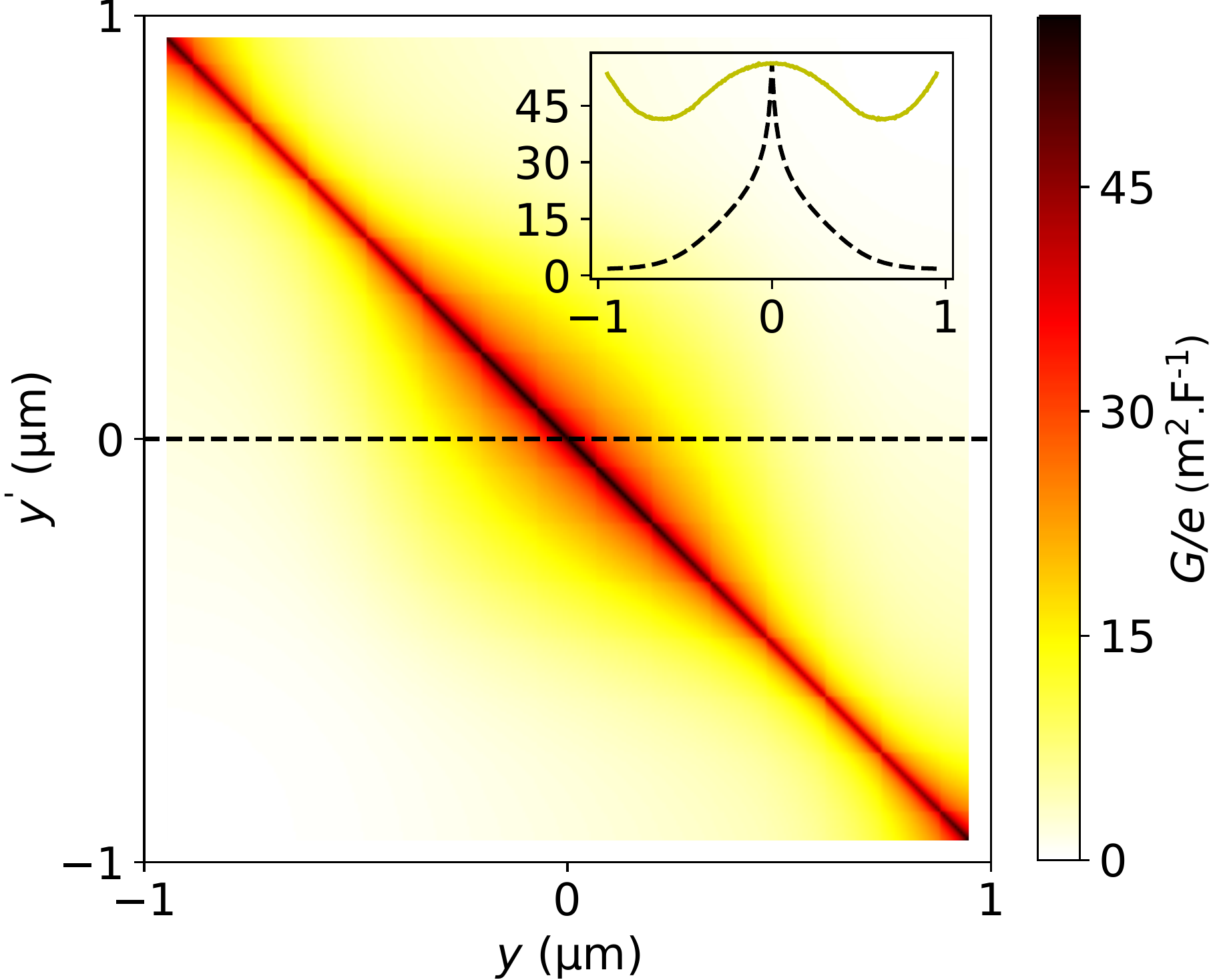}}
	\caption{\textbf{Green's function profile}. The Green's function $G(y, y')$ of the Poisson equation calculated with the finite element method. The two thin regions in the diagonal of the figure at $y,y'=\pm 0.5$ coincide with the positions underneath the electrostatic gate. In the inset, the black dashed line represents a horizontal cut $G(y,y'=0)$ while the solid yellow line is the diagonal part of the matrix $G(y,y)$.}
	\label{fig:green_function} 
\end{figure}	
\noindent
The Schr\"{o}dinger equation is discretized using a simple finite difference scheme and solved using the Kwant package \cite{Kwant_paper}. Solving the sequence of Supplementary Eq. (\ref{equ:2D_poisson}), (\ref{equ:1D_schrodinger}) and (\ref{equ:2D_density}) for an input density $\rho$ results to a new density $\rho^{\textrm{out}}$.
\begin{figure}[h!]
	\centerline{\includegraphics[width=0.6\linewidth]{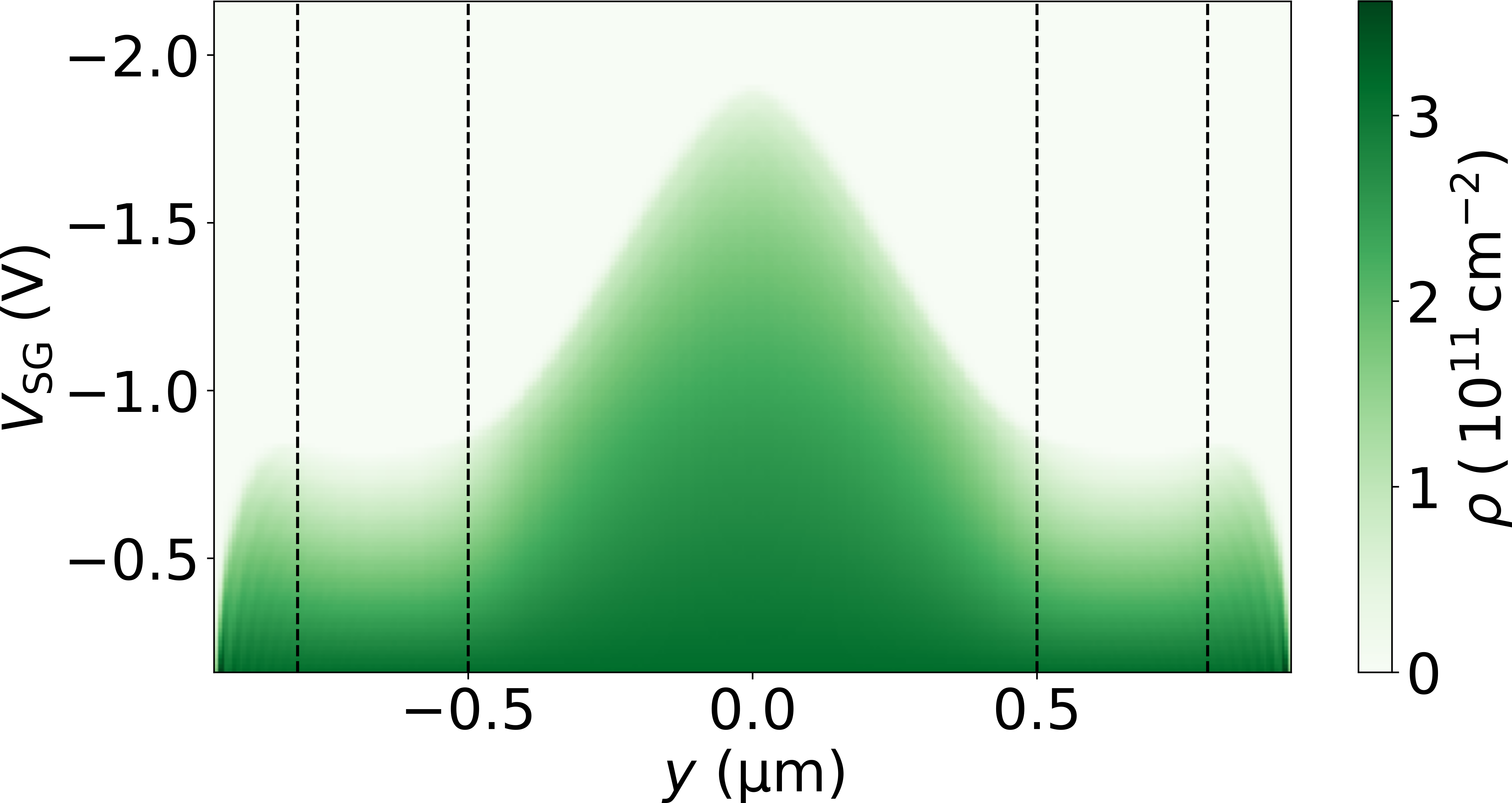}}
	\caption{\textbf{Density profile in the long quasi-1D channel.} Colormap: density as a function of the transverse direction of the wire $y$ (\si{\micro m}) and external gate voltage $V_{\rm SG}$ (V). The vertical black dashed lines show the positions of the two top gates.}
	\label{fig:density_color_plot} 
\end{figure}
The self consistency is reached when $\rho = \rho^{\textrm{out}}$ (see Supplementary Figure \ref{fig:self_consistent_convergence} for an example of convergence of our iterative procedure). The self-consistent solutions are obtained using a Newton-Raphson scheme \cite{Newt_raph}.
We use the following parameters: effective mass $m^\star = 0.067 \, m_e$, dielectric constant $\epsilon = 12 \, \epsilon_0$ and a fixed dopant density of $n_d = \SI{3.16e11}{cm^{-2}}$. Note that this density is higher than the bulk 2D density of the gas.
We have checked that our results are in fact independent of this value since the actual electronic density is controlled by $V_{\rm SG}$. However, using a lower dopant density prevents us from exploring the high density regime ($V_{\rm SG}>\SI{-1}{V}$) where the quasi-1D wire is not defined anymore. 
Supplementary Figure \ref{fig:density_color_plot} shows a color map of the electronic density as a function of the transverse direction of the wire $y$ and $V_{\rm SG}$. The critical values of the gate voltage where
the wire forms ($V_{\rm SG} \sim -0.8 \, \textrm{V}$, the gas is depleted beneath the gates) and the pinch-off ($V_{\rm SG} \sim -1.8 \, \textrm{V}$, full depletion) can be clearly identified.
\begin{figure}[h]
	\centerline{\includegraphics[width=0.6\linewidth]{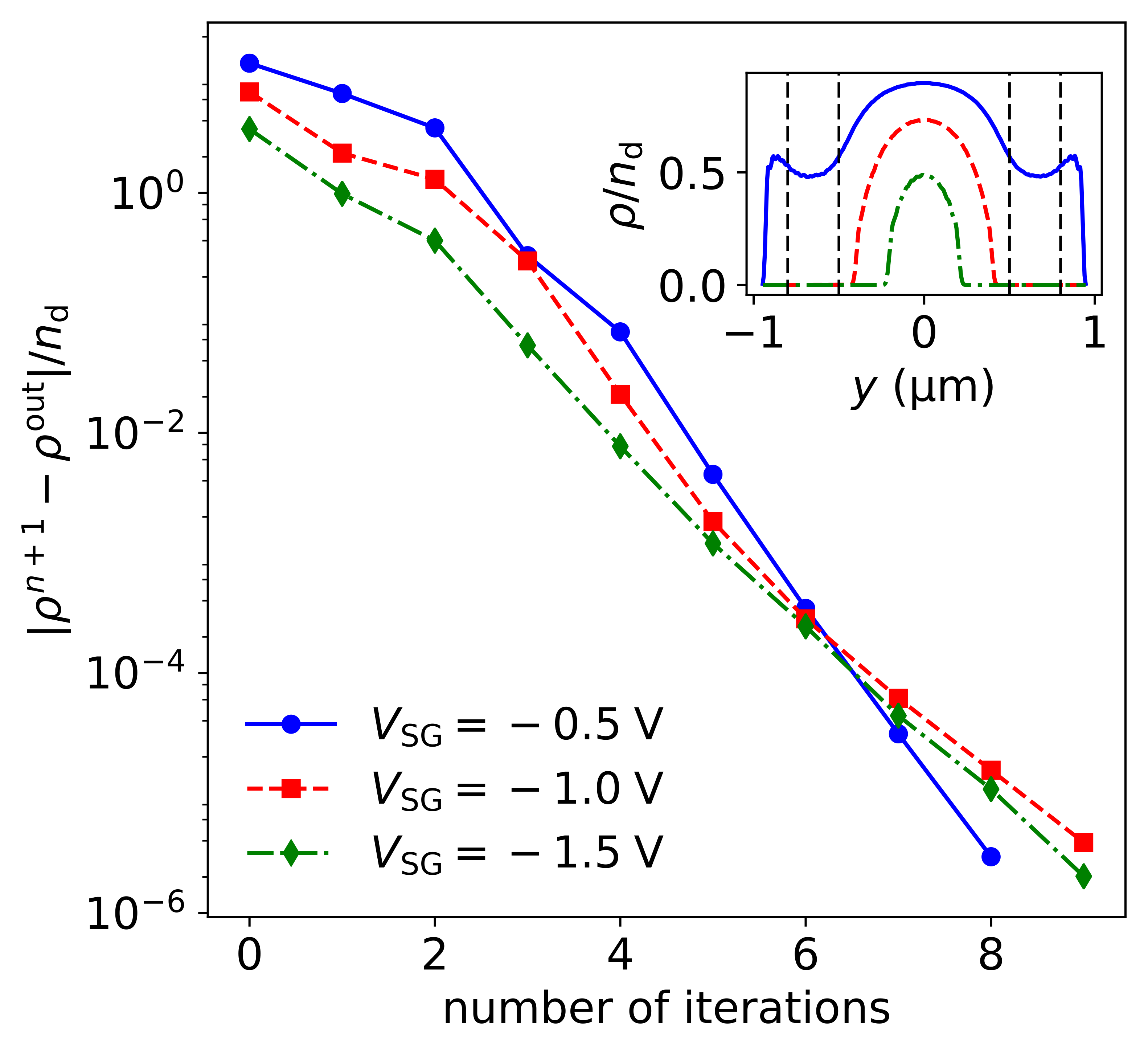}}
	\caption{\textbf{Convergence of our iterative procedure.} Convergence of the self-consistent algorithm for three different gate voltages (solid blue: -0.5 V, dashed red: -1.0 V and dash-dotted green: -1.5 V). At each iteration $n$ we calculate from an input density $\rho^{n}$ a new density $\rho^{n+1}$. Solving sequentially Supplementary Equations (\ref{equ:2D_poisson}, \ref{equ:1D_schrodinger} and \ref{equ:2D_density}) we obtain $\rho^{\rm{out}}$ from which we can calculate the distance to the convergence and express it in terms of the dopant density $n_{\textrm{d}} = \SI{3.16e11}{cm^{-2}}$. The inset is the converged density for the 3 gate voltages in the same units.}
	\label{fig:self_consistent_convergence} 
\end{figure}

\noindent\subsection*{Generalized Luttinger theory} 
\noindent To proceed, we follow Matveev and Glazman \cite{Matveev_glazman} (see also a simpler construction \cite{Kloss}) and construct the bosonised theory for the plasmon excitations of the quasi-1D wire. Bosonization theory predicts that the plasmons have a linear dispersion relation $\omega = v_{\textrm{P}} q$, where $\omega$ is the plasmon energy, $v_{\textrm{P}}$ the plasmon velocity and $q$ the plasmon wave vector. The values of $v_{\textrm{P}}$ are obtained from an eigenvalue problem described below.

\begin{figure}[ht!]
	\includegraphics[width=0.9\linewidth]{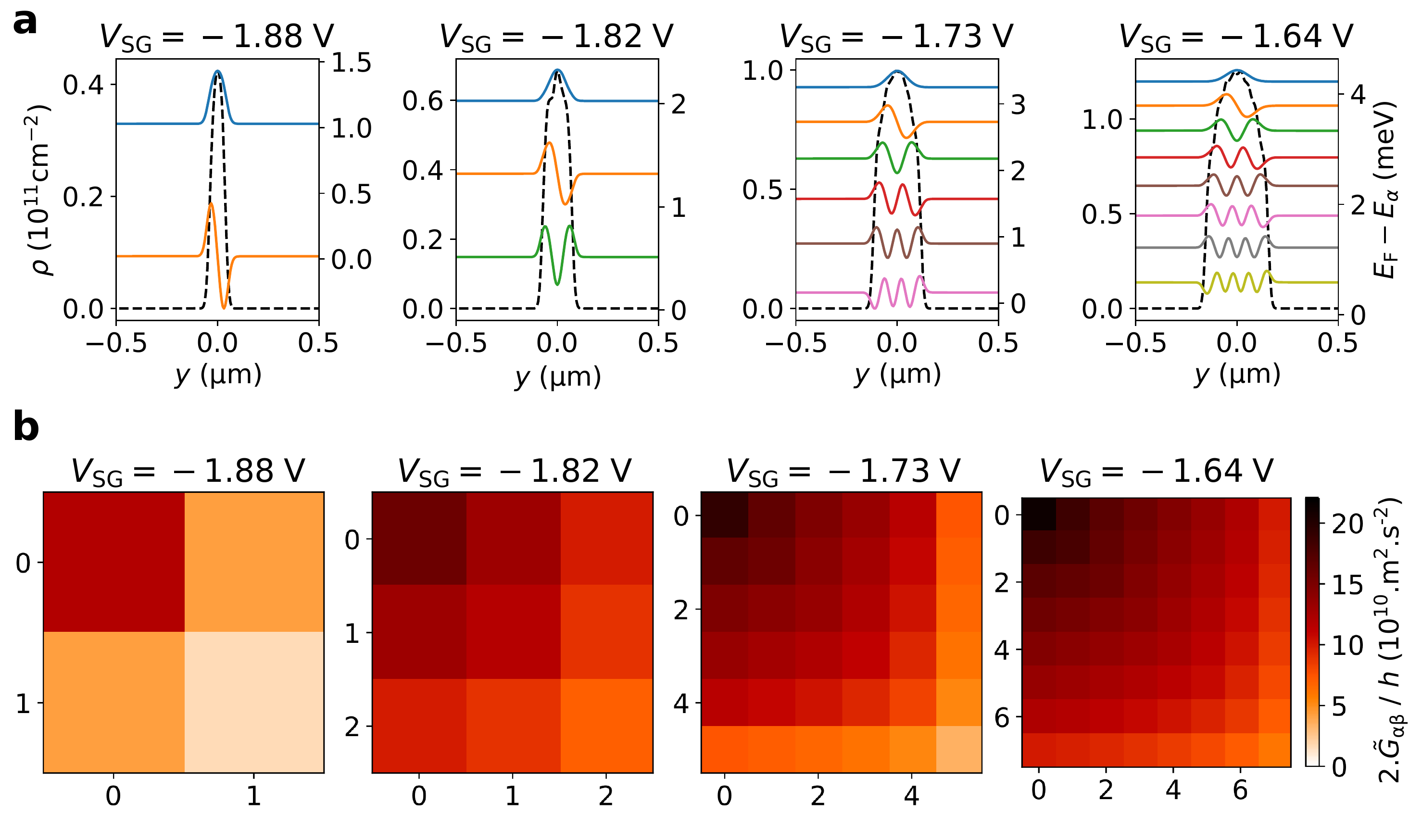}
    \caption{\label{fig:hehe} \textbf{Wave functions in the quasi-1D channel and effective interaction matrix.} \textbf{a}, Density and wave functions for four different channel gate voltages ($-$1.88, $-$1.82, $-$1.73 and $-$1.64 V). In each plot the black dashed curve represents the self-consistent density $\rho$ in cm$^{-2}$ (left y axis). Each coloured line sketches the shape of the wave function (in arbitrary units) of one open mode $\psi_\alpha$, centred around its kinetic energy $E_\textrm{F} - E_\alpha$ (right y axis). \textbf{b}, Green function matrix $\tilde{G}$ from which we calculate the renormalized velocities, Supplementary Eq. (\ref{equ:renorm_v_eigenprob}). }
\end{figure}

\begin{figure}[h!]
	\includegraphics[width=0.9\linewidth]{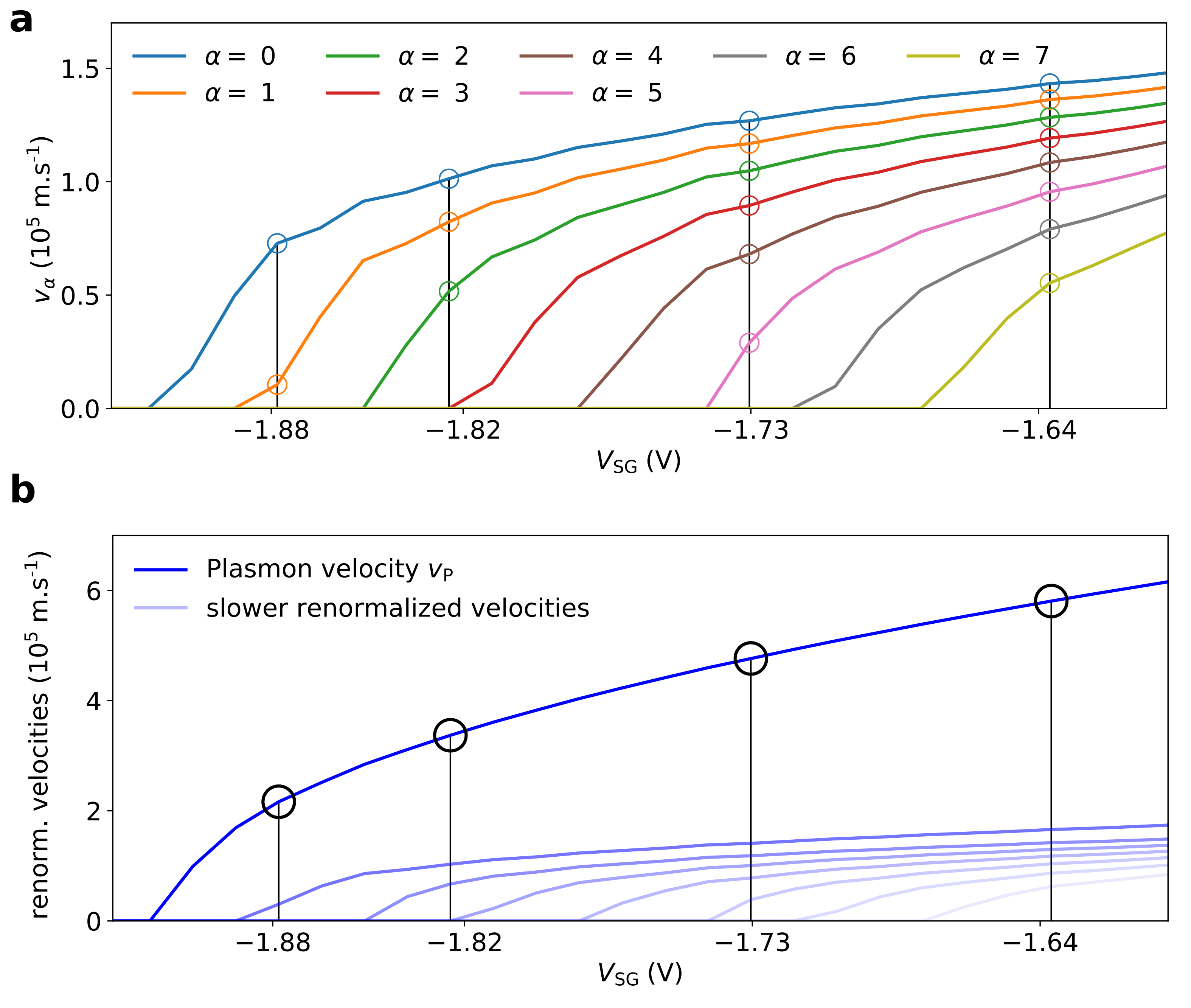}
    \caption{\label{fig:vel_renorm} \textbf{Velocities before and after renormalization.} \textbf{a}, Non-interacting velocities $v_{\rm \alpha}$ for the open channels as a function of gate voltage $V_{\rm{SG}}$. The four gate voltages ($-$1.88, $-$1.82, $-$1.73 and $-$1.64\,V) are marked with black vertical lines and the velocities of the corresponding open modes are marked with circles. \textbf{b}, Renormalized velocities $v_{\textrm{P}}^a$ calculated from Supplementary Eq. (\ref{equ:renorm_v_eigenprob}) for each gate voltage $V_{\textrm{SG}}$. One velocity, indicated by the solid blue line, is considerably higher than the others. This fast collective mode, propagating with velocity $v_{\rm P}$, corresponds to the plasmon mode shown in Fig. 2b of the manuscript.}
\end{figure}

In the presence of $N$ propagating channels, we introduce the
$N\times N$ diagonal velocity matrix $\tilde V$ 
\begin{equation}
\tilde V_{\alpha\beta} = \delta_{\alpha\beta} v_\alpha
\end{equation}
where $v_{\alpha}$ is the non-interacting velocity of mode $\alpha$.
We also introduce the interaction matrix $\tilde G$ defined as,
\begin{equation}
	\label{equ:renorm_g}
	\tilde{G}_{\alpha \beta} = \sqrt{v_\alpha v_\beta} \; \int \, \text{d}y \text{d}y' \;  |\psi_\alpha(y)|^2 G(y, y') |\psi_\beta(y')|^2 
\end{equation}
Supplementary Figure \ref{fig:hehe} and \ref{fig:vel_renorm}a show examples of the different components of $\tilde V$ and $\tilde G$ for different values of the gate voltage. Once these objects have been defined, the plasmon velocities $v_P$ can be obtained in a straightforward manner by diagonalising the following matrix, 
\begin{equation}
	\label{equ:renorm_v_eigenprob}
	\left(\tilde V^2 + \frac{2 }{h} \tilde{G} \right) \bf \tilde{n} =\it v_{\textrm{P}}^{\textrm{2}} \bf \tilde{n}
\end{equation}
where $\bf \tilde n$ is a N-sized vector. Typically Supplementary Eq. (\ref{equ:renorm_v_eigenprob}) has one large eigenvalue -- referred to as the plasmon mode with velocity $v_{\textrm{P}}=v_{\textrm{P}}^0$ -- and $N-1$ small ones (the slow modes) due to the low effective rank of the $\tilde G$ matrix (see Supplementary Fig. \ref{fig:hehe}). The plasmon velocity $v_{\textrm{P}}$ is the chief outcome of this calculation and is shown in Supplementary Fig. \ref{fig:vel_renorm}b. By selecting the $N_\mathrm{ch} \times N_\mathrm{ch}$ submatrix obtained by truncating Supplementary Eq. (\ref{equ:renorm_v_eigenprob}) to the corresponding channels, we reproduce the effect of the channel filtering with QPC$_0$. Diagonalising the $1 \times 1$, the $2 \times 2$ and the full matrix give respectively the red, green and blue curves of Fig. 2b of the manuscript. In the same way, in order to obtain the theory data of Fig. 3d at $V_{\rm SG} = -1$ V, we solve the $N_\mathrm{ch} \times N_\mathrm{ch}$ truncated equation where  
$N_\mathrm{ch}(V_{\rm QPC_0})$ is the number of opened channels of {\rm QPC$_0$}.
%
%

\noindent\subsection*{Modeling of the observed signal} 
The generalized Luttinger theory provides two pieces of information: the velocities $v_P^a$ of the different modes and the eigenfunctions $\tilde{n}_{a\alpha}$ that indicate how mode $a$ decomposes on the different single particle channels $\alpha$. A proper theory of how these modes are generated by the ohmic contact, affected by the presence of the intermediate QPC and eventually measured with the last QPC is beyond the scope of this article. Below, we investigate two limiting cases where we predict the actual shape of the measured signal within a  (i) ``Funneling" scenario and a
(ii) ``Filtering" scenario. In both scenarios, we assume that the Ohmic contact initially populates all single particle states equally, which amounts to assuming that mode $a$ receives an initial weight
$c_a = \sum_\alpha \tilde{n}_{a\alpha}$ (where the global sign of $\tilde{n}_{a\alpha}$ is fixed by imposing $c_a\geq 0$). 

{ \bf{(i) Funneling scenario:}} Within this scenario, the charge hosted by the different channels is funneled into the lowest channel upon entering a QPC (a part can also be reflected, here we are not interested in the absolute height of the signal but in the relative weight of the different modes). At a distance $d$, the expected measured signal takes the form,
\begin{equation}
S(t) \propto \sum_{a=0}^{N-1} c_a f\left(t - \frac{d}{v_{\textrm{P}}^a}\right)
\end{equation}
where $f(t) = \exp(- 4\log (2) t^2/\Gamma^2)$ with a FWHM (Full Width Half Maximum) $\Gamma = \SI{68}{ps}$. Here $f(t)$ corresponds to the Gaussian shape of the injected charge pulse (see Fig. 2a and 5b of the manuscript). 
The corresponding signal is plotted in the upper left panels of Supplementary Figure \ref{fig:signal} a, b and c.

Upon polarizing {\rm QPC$_0$} to a unique transmitting channel $T=1$, one expects that the plasmon is funneled into a plasmon hosted purely by the first channel. In practice this corresponds to truncating Eq.(\ref{equ:renorm_v_eigenprob})
to a $1\times 1$ matrix before diagonalizing the problem. We expect
\begin{equation}
S(t) \propto f\left(t - \frac{d}{v_{\textrm{P}}^{(1\times 1)}}\right)
\end{equation}
where $v_{\textrm{P}}^{(1\times 1)}$ refers to the truncated matrix. The corresponding signal is plotted in the upper panels of Supplementary Figure \ref{fig:signal} a, b and c. The velocity of this pulse corresponds to the red curve shown in Fig. 2b.
A similar calculation with a truncation to a $2\times 2$ matrix leads to the green curve of Fig. 2b. 

{ \bf{(ii) Filtering scenario:}} Upon entering a QPC polarised to transmission $T=1$, only the weight of the mode corresponding to the lowest channel (channel 0) is transmitted, the rest is reflected. This corresponds to adding a factor $\tilde{n}_{a0}$ in the expected signal. The fast plasmon mode has a large charge weight $c_0$ compared to the other modes. However this charge is (more or less equally) distributed over all the single particle channels. In contrast, the other (slow) modes have spread their weight on a few channels with both positive and negative contributions. As a result, the extra factor $\tilde{n}_{a0}$ strongly reduces the overall weight of the fast mode compared to the slow ones. The expected weight without selection {\rm QPC$_0$} reads,
\begin{equation}
S(t) \propto \sum_{a=0}^{N-1} c_a \tilde{n}_{a0} f\left(t - \frac{d}{v_{\textrm{P}}^a}\right)
\end{equation}
while with selection {\rm QPC$_0$} we expect,
\begin{equation}
S(t) \propto \sum_{a=0}^{N-1} c_a (\tilde{n}_{a0})^2f\left(t - \frac{d}{v_{\textrm{P}}^a}\right).
\end{equation}
These signals are plotted respectively on the lower panels of Supplementary Figure \ref{fig:signal} a, b and c for different distances. 


At a short distance (d$\,< \SI{10}{\micro m}$) one observes only a single charge pulse when using the selection {\rm QPC$_0$} which is consistent with both scenarios. For a longer distance ($d\,> \SI{20}{\micro m}$), on the other hand, one observes a clear splitting of the signal into two separate peaks of comparable magnitude (``fractionalisation") for the filtering scenario. This is in contrast to what we see experimentally when measuring the charge pulse at {\rm QPC$_2$} which is placed at a distance of d$\,=\, \SI{25}{\micro m}$ from the selection {\rm QPC$_0$}. 
At this distance we only observe a Gaussian charge pulse with a single peak and which propagates with a much slower speed compared to the case when not using the selection {\rm QPC$_0$}. This lets us conclude that the Funneling scenario is consistent with our experimental data.
For even longer distances one observes complete separation of the fast plasmon mode and the slow modes. The speed of this fast charge mode can be measured when {\rm QPC$_0$} is depolarized and corresponds to the blue curve of Fig. 2b of the main text.

Note that our conclusion stands if we replace the velocities predicted by our modeling by the one actually measured: the fastest velocity that we measure (larger than $\SI{8e5}{m.s^{-1}}$) is sufficiently different from the slow ones (less than $\SI{2e5}{m.s^{-1}}$) for two peaks of FWHM = $\SI{68}{ps}$ show a clear splitting after propagating more than $\SI{25}{\micro m}$.

\newpage
\begin{figure}[h!]
	\includegraphics[width=0.85\linewidth]{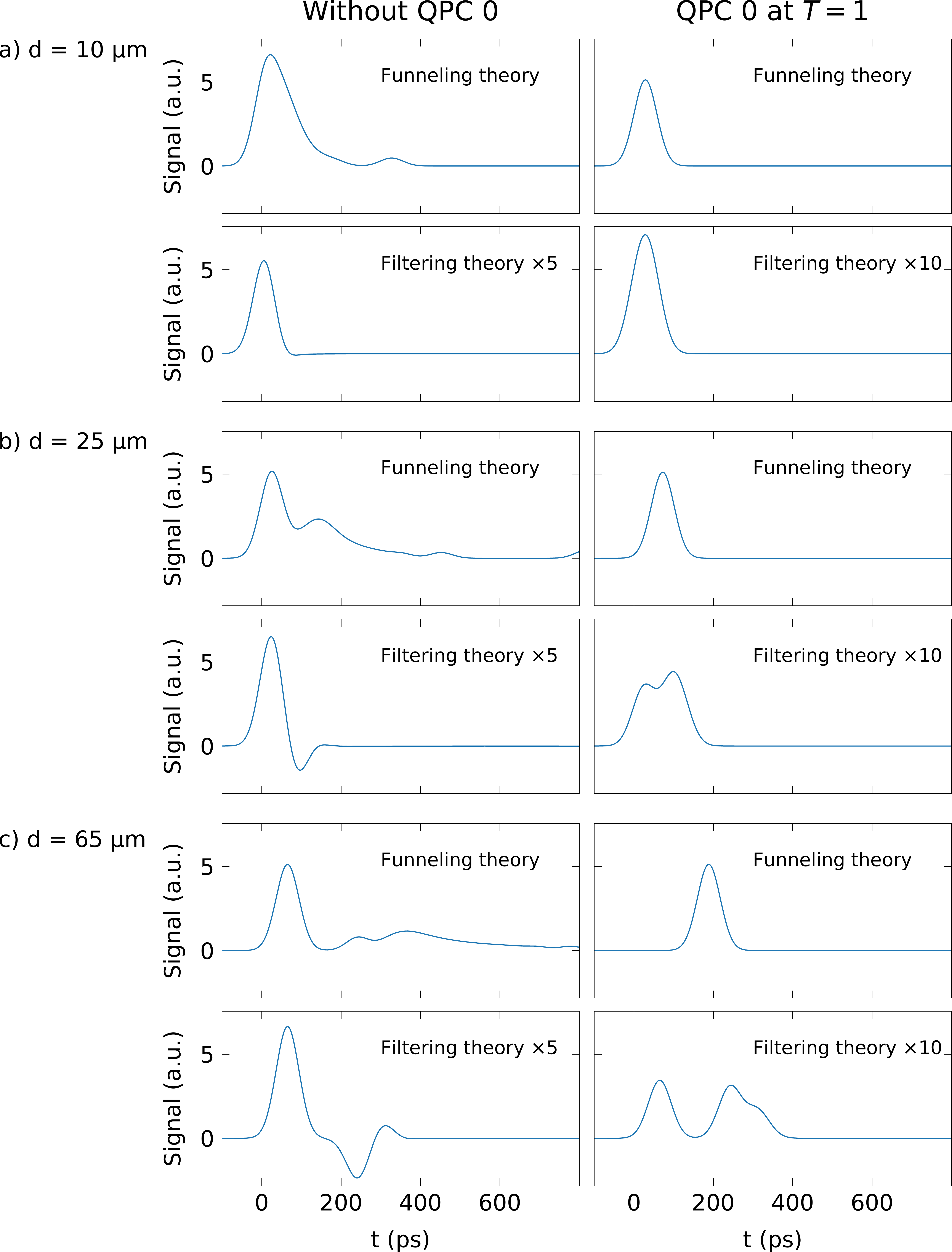}
    \caption{\label{fig:signal} \textbf{Modeling the observed Signal. } Expected signal for the two scenarios (Funneling and Filtering) and two experimental situations (measurement with or without polarising QPC$_0$). The curves correspond to $V_{\rm{SG}}\,=\,\SI{-1.0}{V}$ and a distance of d$\,=\, \SI{10}{\micro m}$ (a), d$\,=\, \SI{25}{\micro m}$ (b) and d$\,=\,\SI{65}{\micro m}$ (c).}
\end{figure}